# Local "hard" and "soft" pinning of 180° domain walls in BaTiO$_3$ probed by *in situ* transmission electron microscopy

Reinis Ignatans, Dragan Damjanovic, and Vasiliki Tileli[*]

Institute of Materials, École Polytechnique Fédérale de Lausanne, 1015 Lausanne, Switzerland

**Abstract**

We report on the electric field response of 180° nanodomain walls in BaTiO$_3$ using *in situ* electrical biasing in transmission electron microscopy (TEM). The sample is biased on a micro-device designed for reliable testing whose key attributes are confirmed by finite element calculations. The presence of weakly charged zig-zag domain walls at room temperature is attributed to the geometric confinement of the device. The motion of the domain walls under the applied electric field allows to extract local P-E loops where distinct domain wall pinning in deep and random energy potential profiles, characteristic for "hard" and "soft" ferroelectrics, respectively, are observed. "Hard" domain wall pinning results in asymmetrical loops typical for "hard" ferroelectrics while the "soft" domain wall pinning follows Rayleigh-like behaviour. All effects are measured locally and directly from the imaged domain structure.

[*]vasiliki.tileli@epfl.ch

# I. INTRODUCTION

Understanding the electrical response, polarization switching, and domain wall movement processes in ferroelectric materials is important for the envisioned future applications such as field effect transistors for neuromorphic systems or the resurgence and optimization of devices including ferroelectric random-access memories (FeRAMs) [1], [2]. Particularly important is polarization switching through nucleation and movement of ferroelectric 180° domain walls as they offer the least amount of emerging strain during operation, thus minimizing issues associated with mechanical failure and cracking [3]. The functionality of many ferroelectric devices depends on the speed of switching of the polarisation states or domains [4], [5] and this is directly linked with the domain wall movement. Understanding the dynamics of the domains walls is highly relevant to already commercially applied ferroelectrics as domain wall movement contributes to the high dielectric permittivity, excellent electromechanical response, and associated electrical, dielectric and mechanical loss mechanisms in ferroelectric perovskite oxides [6], [7].

The origins of ferroelectricity at atomic scale can be understood via various quantum-mechanical, order-disorder and displacive mechanisms [8], [9], [10], [11], [12]. Neglecting the atomistic origins of ferroelectricity, Landau's phenomenological theory and its extensions successfully deal with the phase transitions and the macroscopic physical properties of the material and their dependencies on the external stimuli [13], [14], [15], [16], [17], [18]. Between the atomic and macroscopic scale lays the domain structure of the ferroelectric that can be experimentally accessed via various microscopy and scattering techniques [3]. Dynamic studies of the ferroelectric domain structure under external electric field in the mesoscopic length-scale is relevant due to the strong influence on the macroscopic properties of the material.

The domain response to the electric field is typically probed indirectly using dielectric spectroscopy, polarization- or strain- response as a function of field, current response during switching or *in situ* diffraction techniques [19], [20]. These methods reveal mostly the average behaviour of the system. Studies of the local domain response usually involve piezo-force microscopy (PFM) or *in situ* transmission electron microscopy (TEM) [21], [22], [23]. With PFM quantitative measurements are difficult and the electric field is inherently inhomogeneous [24]. Technical requirements for interpretable *in situ* electric-field biasing results in the TEM also make the experiments challenging. For instance, specimen preparation is required to result in a sample with high electrical resistivity otherwise the current would flow across the conductive channels and the actual electric field would be smaller than the nominal (applied) field. In that case, the observed response would be dominated by Joule heating instead of the applied electric field. In most cases localized Joule heating leads to sample failure (cracking, electrode delamination due to the current surge, or complete sample disintegration). In all cases, it is necessary for the applied electric field to be homogeneous across the probed area in order to reduce complexity in interpretation of the observed dynamic processes.

To perform *in situ* electrical biasing experiments in the TEM, several methods are available. Traditionally, biasing experiments were performed on TEM holders that could accommodate mechanically polished samples connected with electrodes far away from the area of interest, which could be contacted in several ways (wire bonding, silver paste etc.) [23]. The main advantage of this method is the excellent TEM sample quality, [23], [25], [26], [27], [28] however, the resulting electric field in a mechanically polished sample cannot be uniform and it is also difficult to control the geometry of the region of interest with respect to the direction of the electric field [28], [29]. For better control of the applied voltage, the probe technique is used for biasing ferroelectric structures in the TEM. In this case, a metallic probe is pressed against the ferroelectric material to apply the potential [30], [31], [32]. However, since ferroelectrics are highly responsive on the applied external stress, pressing the probe against the material can adversely influence the formation and switching of the domains. Additionally, this method creates a highly inhomogeneous local electrical field, which hinders interpretation of the results. These issues can be mitigated by sandwiching the ferroelectric between conductive layers and pressing the probe on the conductor far away from the region of observation [33]. Nevertheless, such configuration is viable for thin films [34], but not for samples with larger areas such as single crystals or ceramics and restrictions imposed on the sample geometry with the addition of conductive layers can influence the inherent switching properties of the ferroelectrics through interfacial effects. More recently, advances in TEM holder instrumentation enabled microelectromechanical (MEMS) Si-based chips with patterned electrodes deposited on silicon nitride membranes to be used for heating and biasing experiments [23]. The sample is transferred on such chips using standard focused ion beam (FIB) lamella-making protocols [35]. The advantages of this method involve the ability to control the sample orientation and, consequently, to control the desired electric field direction with respect to the crystallographic orientation. However, ferroelectric sample preparation is especially challenging due to possible strain fields and emerging conduction channels due to ion implantation, surface amorphization and, redeposition of residuals while polishing the lamella [36], [37], [38].

Herein, we used MEMS-based microchip TEM holder technology to systematically bias a single crystal $BaTiO_3$ sample along the pseudo-cubic $[100]_{PC}$ direction. In particular, we use an optimized sample geometry that enables us to precisely control the applied voltage and, in consequence, the electric field distribution in the probed area. The electrical performance of the device is confirmed with simulations using finite element methods. To quantitatively evaluate our results, we focused on the motion of 180º domain walls at room temperature.

Local physical phenomena such as weakly charged zig-zag domain walls and domain wall pinning as a function of the applied voltage are directly probed.

## II. METHODS

### A. Sample preparation

Single crystalline BaTiO$_3$ (001) (MTI corporation) was used for the device fabrication. The largest facets of the crystal were already polished to optical grade. For the preparation of the lamella, the top surface was sputter-coated with 30 nm thick carbon layer (Cressington 108 series carbon coater). This step is required to dissipate the charges during electron imaging and to protect the surface from Ga ion implantation during ion beam milling.

Inside the focused ion beam (FIB) milling instrument (Zeiss NVision 40, Ga$^+$ source), two site-specific carbon protective layers were additionally deposited. A thin electron-beam deposited (thickness in the range of ~100 nm at 5 kV) followed by a thicker ion-beam deposited protective layer (1.5 μm thickness at 30 kV and 150 pA current). Trenches around the protected area were etched consecutively with 27, 13, 6.5, and 0.7 nA ion beam currents at 30 kV. The lamella was thinned to about 1 μm prior to its transfer with a micromanipulator (Kleindiek Nanotechnik) to a 4-heating and 2-biasing MEMS chip (DENSsolutions). Complete details of the final steps of the device fabrication and the measurement of the thickness of the final lamella can be found in Supplemental Material, SM [39].

Finally, the temperature on the heating element of the heating/biasing TEM holder was pre-calibrated using its electrical resistance and the electrical bias to the sample was applied with a source meter (Keithley SMU-2450) using a script to generate a triangular voltage waveshape.

## B. Finite element calculations

The electrostatic finite element model (FEM) of the device was created with COMSOL Multiphysics 5.4 package. Dimensions for the electrostatic model were measured from scanning electron microscopy (SEM) images. The dimensions of the simulation box were 40 μm width, 30 μm height, and 30 μm in depth. Tetrahedral adaptive mesh was used for the simulation. The general mesh element characteristics for the simulations involved the electron transparent window of 0.02 μm max and 0.008 μm min, the bulk part of the lamella of 0.05 μm max and 0.008 μm min, the surface of the electrodes of 2.2 μm max and 0.16 μm min, the SiNx membrane of 0.2 μm max and 0.008 μm min, and the vacuum region around the device 0.25 μm max and 0.008 μm min. The bulk part of the lamella was set to 1 μm thick and the electron transparent region to a thickness of 268 nm. The potential difference between the electrodes in the calculations was set to 1 V and it was used as a boundary condition.

## C. Electron Microscopy

Bright field scanning transmission electron microscopy (BF-STEM) was performed on a double spherical aberration (Cs) corrected Titan Themis 60-300 (ThermoFischer Scientific) operated at 300 kV. A beam current of 100 pA was used, the beam convergence angle was 28 mrad and the collection angle was 79 mrad. Serial STEM imaging was done with a pixel scan area of 1024x1024 (3.204 nm/pixel) and 2.0 μs dwell time resulting in 2.98 s/frame. The images were then cropped to 764x764 or 760x1024 pixel size. Images were cropped to remove completely dark parts from the sides of the electron transparent window. Post-processing was performed using ImageJ software to enhance the domain wall contrast. It involved the following steps: Fourier filtering to reduce the horizontal scan noise from the images, subtraction of the background (50 pixel rolling ball radius), and contrast and brightness adjustment.

## III. RESULTS AND DISCUSSION

We used a six-contact MEMS chip (Fig. 1(a)) that utilizes four of the contacts for heating and precise temperature control and two for biasing (Fig. 1(b)). The BaTiO$_3$ lamella was FIB-prepared and planar Pt contacts were ion-deposited on the two sides of the lamella as depicted by the SEM image in Fig. 1(c). The sample was then thinned to a final thickness of 268 nm (details on the electron energy-loss spectroscopy-based thickness measurement are provided in SM and the thickness map is shown in Fig. S1 [39]). As a final step, to mitigate conductive paths caused by the contaminants introduced by the ion-beam preparation method, site specific etching was performed around the area of interest (green boxes in Fig. 1(c)).

Fig. 1(d) shows FEM calculations of the field distribution for the final geometry of the device. The resulting electric field, E, in the central part of lamella is calculated to be ~1.65 kV/cm for 1 V applied to the electrodes. This value is the upper bound of the electric field magnitude since the simulation does not consider leakage currents and effects of the interfaces. These effects might alter the magnitude of the applied field, however, the overall characteristics of the electric field distribution remain the same. The calculations predict a difference in the electric field of ~0.4 kV/cm field across the vertical direction of the lamella (white arrow in the Fig. 1(d)). Fig. 1(f) depicts the magnitude of the electric field along the arrow line and its dependence on the dielectric constant of the sample). The inhomogeneity of the electric field and small gradient across the lamella results from the etched-out regions. This inhomogeneity can be reduced by more conservative isolating cuts, however, this increases the probability of failure of the device during biasing. Considering the geometry of the lamella, for a given applied voltage the magnitude of the electric field in the sample is mainly influenced by the width (horizontal white arrows in Fig. 1(d)) and thickness of probed area. If the width of the probed area decreases (Fig. 1(e)), the magnitude of the electric field increases, which is similar to decreasing the plate distance in a parallel plate capacitor configuration. Reduction of the thickness of the lamella would also result in an increase in the magnitude of the applicable electric field (Fig. 1(e)) as the equipotential lines get "squeezed" together in the smaller volume. The overall electric field distribution is also dependent on the dielectric permittivity of the sample. Fig. 1(f) shows the dependence of E on the average dielectric constant of the probed material along the central part of the lamella (white vertical arrow in the Fig. 1(d)). For low values of the relative dielectric permittivity (below 100), the resulting electric field is highly dependent on the permittivity of the material. When the dielectric permittivity approaches the values corresponding to perovskite ferroelectrics (dielectric permittivity ranging between hundred to several thousands), the dependence of the field on the permittivity becomes minimal. Thus, the dielectric properties of $BaTiO_3$ do not influence the magnitude of the electric field along the probed area.

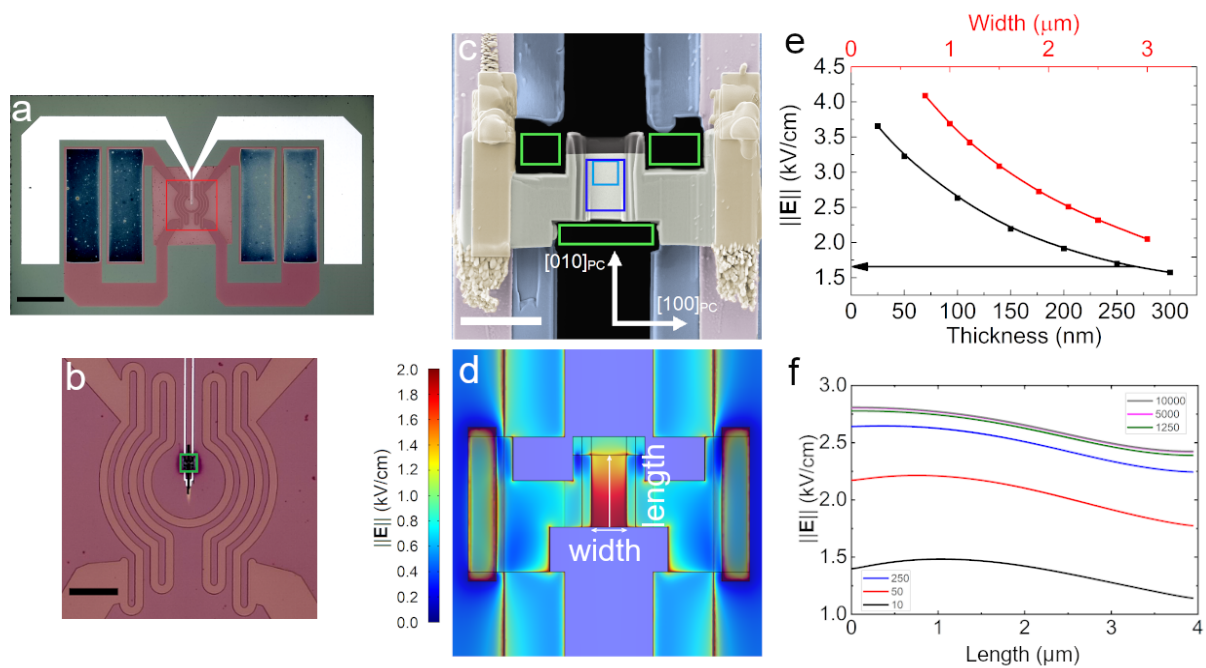

FIG 1. (a) Optical microscope image of the in situ biasing/heating TEM MEMS chip. Scale bar is 500 μm and red rectangle corresponds to the area depicted in (b). (b) SiN$_x$ membrane region with the heating spiral and two biasing electrodes (bright vertical features). Scale bar is 100 μm and small green rectangle corresponds to the location of the lamella. (c) False colour SEM image of BaTiO$_3$ sample mounted on the MEMS chip (black area corresponds to the vacuum region, blue to the silicon nitride membrane, pink to the Pt electrodes on the membrane, yellow to the FIB deposited Pt contact electrodes, grey to the bulk part of BaTiO$_3$, light grey to the electron transparent window, and dark grey to the protective carbon layer). Scale bar is 5 μm. Green boxes represent the etched-out areas, dark blue rectangle represents the area for domain area – voltage loops determination and light square shows area imaged in Fig. 3. (d) Plot of the electric field distribution within the device as calculated by finite element methods, white arrows indicate width and length directions. (e) Plot showing the dependence of the calculated electric field on the thickness and width of the electron transparent area ($\varepsilon$ = 1250, for varying thickness, the width was fixed at 2.275 μm; for varying width, the thickness was fixed at 100 nm), black arrow shows the calculated electric field for the actual lamella thickness (268 nm). (f) Plot showing the dependence of the calculated electric field on the lamella's dielectric constant. In both (e) and (f) the electric field is plotted along the central part of the electron transparent region (white vertical arrow in the (d)).

Prior to biasing the sample, annealing above the Curie temperature ($T_c$) was performed inside the TEM. Heating is required to release most of the strain due to the FIB preparation process [36]. The sample was heated to 200 ºC, held at this temperature for 1 minute, then the temperature was lowered to 25 ºC in 20 min (the heating profile and respective image sequence are shown in Supplemental Video 1, SM [39]). The biasing experiments were performed at room temperature where the sample is dominated by unusually stable 180º walls [40]. Fig. 2(a) depicts bright field scanning TEM (BF-STEM) images of the domain structure at room temperature (the full image sequence during electrical biasing is shown in Supplemental Video 2, SM [39]). The dark contrast in the images corresponds to the domain walls. To improve this contrast without interference due to diffraction effects, the sample was mistitled by ~10 mrad from the [001]$_{PC}$ zone axis and was postprocessed. The sample was then biased with a triangular waveform ($\Delta U$ = 0.0351 V/s in the case of $U_{max}$ = 3.5 V, while $\Delta U$ was reduced for smaller $U_{max}$ to improve temporal resolution) starting from the positive direction to the maximum of the positive voltage (3.5 V, Fig. 2(c)), continuing to negative applied voltage (-3.5 V, Fig. 2(g)) before completing the loop at 0 V (Fig. 2(i)). We note that the applied voltage of 3.5 V corresponds to electric field values of about 5.8 kV/cm, based on the FEM simulation. Considering the coercive field of BaTiO$_3$ (approximately 0.5 kV/cm), by studying the domain response we conclude that the actual applied field is within one order of magnitude of the FEM calculated one [41]. This discrepancy is attributed to the sample-electrode interface and leakage current effects, which are difficult to be eliminated due to the microscopic size of the device and which are not included in the calculations of the electric field (more information on the equivalent expected electric circuit of the device, Fig. S2, and on the discrepancy of the FEM-calculated electric field can be found in SM [39]).

By examining the domain structure in the BF-STEM images, we firstly note the presence of weakly charged zig-zag domain walls (Fig. 2(a)). It is theoretically predicted that weakly charged domain walls are more stable in BaTiO$_3$ when it is not in a bulk form and it is connected with platinum electrodes [42]. This corresponds well to the geometry of our device where the polarisation is essentially confined in a thin slab within two dimensions with Pt electrodes on the sides. The transition mechanism of the strongly charged domain wall into weakly charged zig-zag domain wall and the meaning of the periodicity has been previously described by Sidorkin [43].

The polarisation direction within the domains can be determined upon application of the electric field (Fig. 2(c)). The domain with polarisation pointing in the direction of the electric field should grow and, consequently, domains with polarisation vector opposite to the applied field should shrink. By examining domain and domain wall response on positive bias (the direction of the applied electric field is depicted with a black arrow at the start or end of the image rows in Fig. 2), the polarisation vectors in the visible 180º domains were determined and are indicated with red arrows in Fig. 2(c). Consequently, red shading in the BF-STEM images of Fig. 2 corresponds to domains with polarisation direction pointing to the right. It is noted that the periodic vertically aligned 180° domain walls, which appear as weak contrast in the central part of Fig. 2(a-f), do not respond as strongly to the external bias since the polarisation and the electric field vectors are orthogonal.

When the direction of the electric field is changed, Fig. 2(d-g), the domains that span across the whole electron transparent window as bands, experience sideways growth. The observed domain wall bending in Fig. 2(g) is attributed to strong domain wall pinning possibly due to a dislocation centre. The domain wall displacement, $u(x)$, from the equilibrium position (which in this case is aligned along the $[100]_{PC}$ direction) has characteristic hyperbolic nature (theoretically $u(x) \sim \frac{1}{\sqrt{x}}$), as the domain wall regains equilibrium when the distance, $(x)$, from the pinning centre increases [43].

Finally, the applied bias cycle terminates after completing the full loop (Fig. 2(i)). During the field cycling the domain structure exhibits hysteretic behaviour at all fields by not returning to the same state for the same voltage during increasing and decreasing steps of the full cycle (for example, compare panels b and d for +1.75V and a, e, and i for 0 V, in Fig. 2).

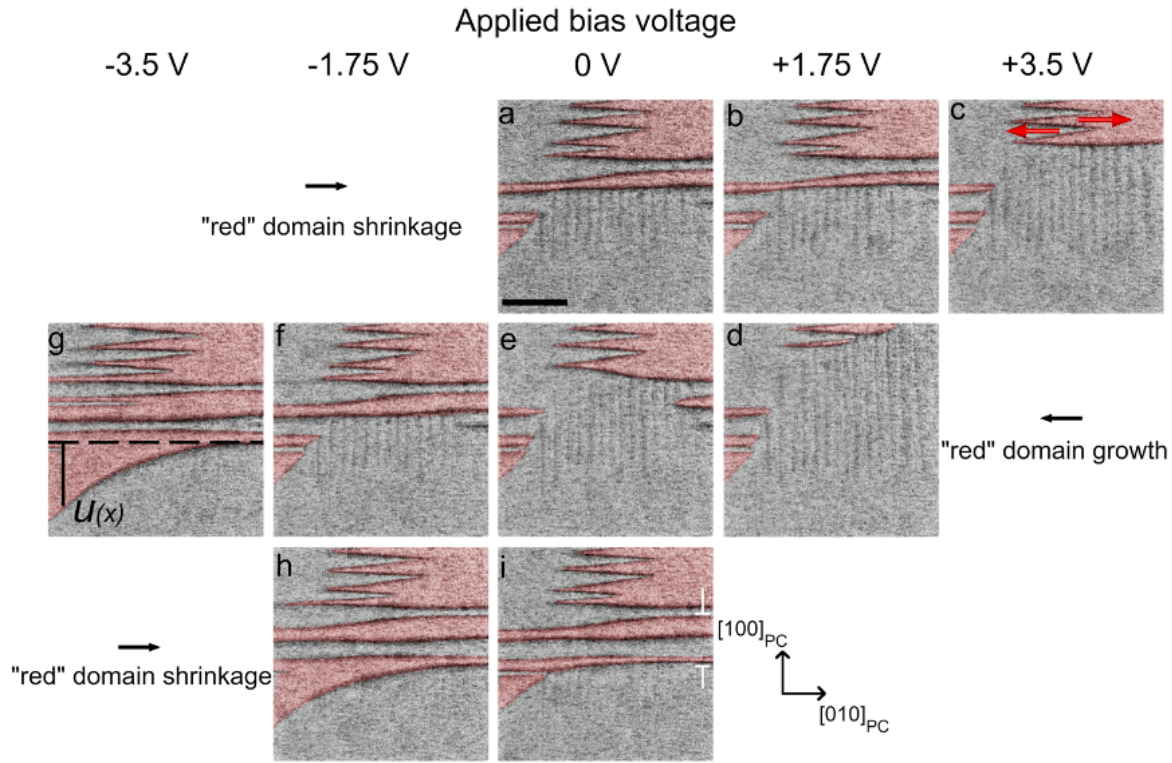

FIG 2. BF-STEM micrographs showing the domain structure evolution during cyclic electrical biasing. In the image, the field is oriented along the $[010]_{PC}$ and black arrows indicate the direction of the electric field. (a) Initial domain structure at 0 V bias. (b) Domain structure at +1.75 V bias with increasing electric field strength. (c) Domain structure at maximum of +3.5 V bias. Red arrows indicate the polarization direction in the two domains (d) Domain structure at 1.75 V with decreasing field strength. (e) Domain structure at 0 V in the middle of the cycle. (f) Domain structure at -1.75 V with increasing field strength. (g) Domain structure at minimal voltage -3.5 V. u(x) represents the magnitude of domain wall bending from the equilibrium position, which is marked with the dashed line (g). (h) Domain structure at -1.75 V with decreasing field strength. (i) Final domain configuration at the end of the cycle with 0 V. White indication lines depict the double domain that shows sideways growth and it is hysteric to the domain in (a). Red false-colour superimposed on all images corresponds to the domains with polarisation pointing to the right. Scale bar is 500 nm (shown in (a)).

The response of the domains at different maximum applied voltages was further investigated to study the domain dynamics from weak to strong fields. The voltages ranged from 0.25 to 3.5 V for 14 different values in total and all image sequences are shown in Supplemental Videos 3-16, SM [39]). Analysis of the sequential BF-STEM images of the domains exhibiting the same polarization (i.e. the red shaded domains in Fig. 2), resulted in the domain area vs. applied potential loops shown in Fig. 3(a) (each data point corresponds to the domain structure image at given bias potential plotted as bias voltage–domain area loops for three cases corresponding to 3.5, 2.5, and 2.0 V maximum applied voltage, the methodology of calculating the loop at 3.5 V is detailed in Fig. S3 in SM [39]). The measured area is proportional to the polarisation pointing to the right and this corresponds to domain growth with respect to negative applied voltage. The measured loops in Fig. 3(a) resemble the polarisation-electric field (P-E) loops in "hard" ferroelectrics where P-E loops are constricted at weak fields and open at larger fields [44], [45], [46], [47]. Remarkably, the local loops shown here reveal additional features which cannot be discerned in loops taken with classical approaches over macroscopic areas. These include domain wall pinning (weakly responsive polarization region in the -1 V to 1 V interval of the measured loops) and domain nucleation/annihilation accompanied with ballistic movement of domain walls. The apparent loop asymmetry and the fact that at 0 V the relative domain coverage does not reach 50 % (which would be a requirement for electro-neutrality) can be partly attributed to the nature of the local TEM measurements that do not necessarily correspond to the entire domain structure. On the other hand, the loop asymmetry and apparent pinching around 0 V bias indicate hard pinning of domain walls, which are released from pinning centres at higher fields as indicated by the loop opening. These defects exert restoring forces on the domain walls at weak fields limiting the movement of the wall and consequently pinching the loop. At higher electrical fields, the restoring forces are overcome and the domain wall movement is controlled by pinning centres randomly distributed in space and with variable strengths [47], [48].

Fig. 3(b) depicts the totality of the results from the loop biasing experiments plotted as the minimal and maximal cycle voltage for each loop as a function of the domain surface coverage (again, the domains corresponding to the red shading in Fig. 2 were measured). At low voltages, the domain coverage reveals that new domains do not nucleate and existing domain walls do not move significantly as expected for a hard domain wall pinning (this region is marked with orange shading in Fig. 3(b)). At higher applied voltages (above ±1 V), the existing domain walls experience significant forward and sideways motion and domains start to nucleate or annihilate depending on the applied voltage as previously predicted [3]. Overall, a quadratic behaviour of the area (~polarization) vs electric field relation is observed, particularly pronounced at positive applied voltages. The function $P \sim E^2$, which essentially corresponds to classical Rayleigh's law, represents the weak domain wall pinning on randomly distributed defects within the crystal [49], [50]. In principle, Rayleigh's law is applicable at weak field conditions in a material with a broad distribution of pinning centre strengths, but in our case the local domain wall movements are hindered by hard pinning at weak fields, therefore domain walls do not move substantially and do not experience soft pinning upon random lattice defects. Rayleigh motion is achieved only at higher fields when the domain wall is de-pinned and can move through the lattice with randomly distributed defects. The asymmetry of the loops indicates asymmetrical distribution of defects in the examined local area, which is again a feature that would not be necessarily noticed in a macroscopic loop measured over a much larger area. Similarly, the different slope of the positive and negative applied voltages of Fig. 3(c) indicates the different response of the domain walls upon reversal of the magnitude of the electric field.

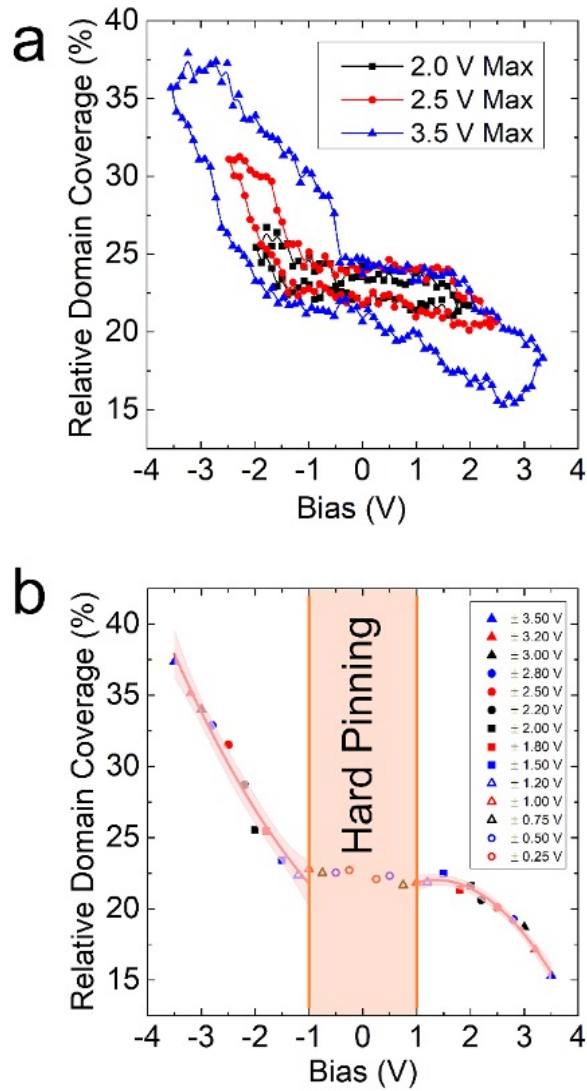

FIG 3. (a) Selected domain area-voltage loops using triangle wave voltages of 2, 2.5, and 3.5 V. (b) Relative domain coverage with respect to the maximal/minimal cycle voltage for the positive and negative branch of the loops (14 loops in total). The lines represent the quadratic polynomial fits for each branch and red/grey shaded regions correspond to 95% confidence interval. Orange shading corresponds to the flat region of the plot, where hard domain wall pinning occurs.

**IV. CONCLUSIONS**

In summary, we demonstrated a specialised sample preparation method, which allows to perform reliable *in situ* electrical biasing measurements in the TEM and provides the ability for interpretable physical processes of ferroelectric nanodomains. The meticulously designed and fabricated geometry of the device results in a homogeneous electric field, whose magnitude and direction are confirmed by finite element calculations. We used the classical ferroelectric $BaTiO_3$ and we showed that precise control of the external electric field and sequential BF-STEM imaging permits to locally study 180° domains and domain wall movements at room temperature. The origins and stability of weakly charged zig-zag nanodomain walls are associated with the thin slab geometry of the ferroelectric and interplay of $BaTiO_3$ with Pt electrodes. Polarization-electric field loops calculated directly from the areas of the imaged domains elucidate the relation between bulk and nanoscale ferroelectric effects. The domain wall response at low electric field is associated with the hard pinning mechanisms on the defects within the lattice, which results in pinched and asymmetric domain area vs. voltage loops. At higher fields, domain walls are de-pined and localised effects such as new domain nucleation and sideways growth and domain wall bending were observed. At such fields (above the coercive field), we show that the domain wall motion follows Rayleigh–like behaviour. Thus, this controlled electrical biasing method can provide unprecedented insights on dynamic domain wall motion and domain interactions at the nanoscale

## ACKNOWLEDGEMENTS

The work was supported by the Swiss National Science Foundation (SNSF) under award no. 200021_175711. D.D. acknowledges support form ONR-Global (award no. N62909-18-1-2078).

# Supplemental Material

# Local "hard" and "soft" pinning of 180° domain walls in BaTiO$_3$ probed by *in situ* transmission electron microscopy

Reinis Ignatans, Dragan Damjanovic, and Vasiliki Tileli

Institute of Materials, École Polytechnique Fédérale de Lausanne, 1015 Lausanne, Switzerland

**Device fabrication**

Before transferring the semi-prepared lamella (~ 1 µm thick, 20 µm wide, 7 µm height) to the MEMS chip, the thin SiN$_x$ membrane in the window region was etched creating a hole and extending this hole along the direction of the electrodes (as seen in Fig. 1b-c of the main manuscript). This modification ensures that the contaminants from the FIB-deposited Pt contact electrodes and the residuals of the final polishing steps do not accumulate in the region between the electrodes, but rather fall in the void. Thus, eliminating the possibility of short circuit due to redeposition layers on the chip.

After transfer, the Pt contacts were deposited firstly underneath the lamella to secure it to the chip at ~9° angle. This ensures good enough visibility upon ion polishing. Then additional Pt layers were deposited on the front side of the lamella (as seen in the Fig. 1c of the main text). Therefore, creating wide planar contacts on the sides (important pre-requisite to achieve homogeneous electric field in the electron transparent window). Pt contacts were deposited with an ion beam voltage of 30 kV and 40 pA current.

The final polishing of the lamella took place at 0.7° relative to the ion beam at 30 kV and 40 pA. Final cleaning of the sides of the lamella was performed at 5 kV and 30 pA with the polishing angle of 5° from both sides.

Additionally, three areas were further etched from the specimen. Two incisions were made at the top part and one large area was cut off at the bottom part of the lamella (the three cuts can be seen in Fig. 1(c) of the main manuscript). The top cuts are made to disconnect the protective carbon from the electrodes. The undercut incision was made after the final polish of the lamella, as this region is prone to accumulate polishing residuals, which often create short circuit across the bottom part of the electron transparent region. The isolating incisions are made with the ion beam normal to the surface of the lamella. Such geometry greatly reduces the chance of redeposition. Care must be taken not to expose the electron transparent window during this operation, therefore only the parts which were etched were exposed to the ion beam. The two isolating incisions in the protective carbon region (top) were made with 30 kV and 150 pA and undercut region (bottom) was etched with 30 kV and 80 pA.

**Measurement of thickness of the lamella**

The thickness of the lamella was measured using low loss electron energy-loss spectroscopy (EELS) on the same instrument (ThermoFischer Scientific Themis 60-300). An electron-beam current of 100 pA was used. Operating conditions included 70 µm C2 aperture, beam convergence angle of 28 mrad, and collection angle of 19.8 mrad.

First, a controlled experiment was performed. A BaTiO$_3$ lamella sample with known thickness was spectrally mapped at the low-loss energy region. The spectrometer dispersion was set to 0.25 eV/channel. From the acquired map, the $t/\lambda$ calculation was done (where $t$ is the thickness of the lamella and $\lambda$ is the mean free path of the electrons in the specimen). The thickness of the lamella was measured with the SEM and the mean free path, at the specific conditions, was calculated to be $\lambda = 156 \pm 3$ nm.

Second, the sample used in biasing experiments was mapped at the same conditions, giving $t/\lambda$ measurements across the whole lamella and, by using the mean free path calculated from the control sample, the thickness map of the imaged area during biasing was measured, Fig. S1. The exact value is calculated to be 268 nm.

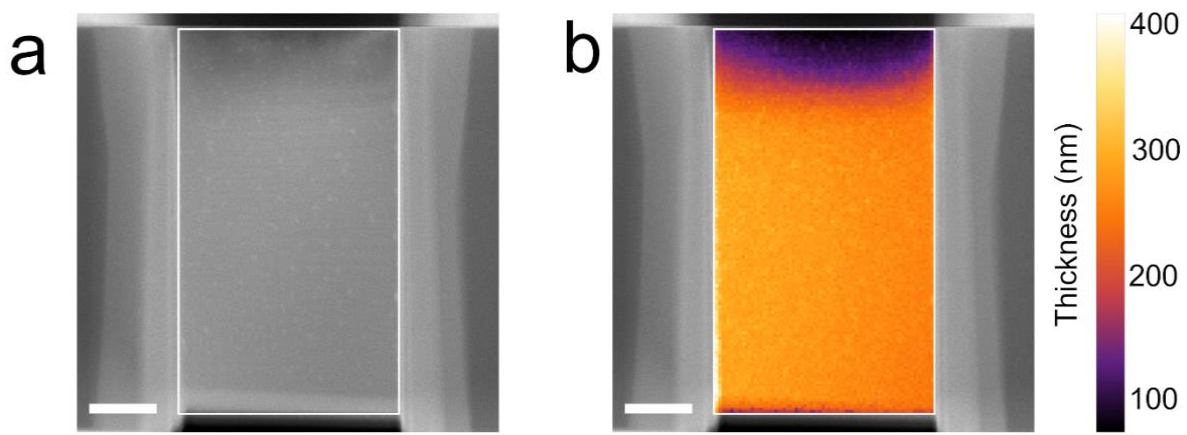

FIG. S1. Thickness map of BaTiO$_3$ lamella used for the biasing experiments. (a) – Annular dark field (ADF) STEM image annotated with a white rectangle which represents the low-loss EELS mapping area. (b) ADF image with overlayed thickness map. Scale bar in both images is 750 µm.

**Considerations of electric field calibration**

A schematic illustrating an equivalent circuit of the microdevice fabricated for the biasing experiments is used to comprehend the expected electron transport behavior across the biased $BaTiO_3$ sample and explain the discrepancies between the applied voltage and expected electric field as calculated with finite element methods. R1 is the electric resistance of the wires leading to the chip in addition with the resistance of the platinum electric lines of the MEMS chip - R1 = 40 Ω. Part of the system consisting of R2, R3 and C is equivalent to the lamella with FIB deposited contacting layers. R2 is the electric resistance of the deposited contact layers between the MEMS Pt lines and the $BaTiO_3$ lamella. R3 is the resistance of the conductive channels within the $BaTiO_3$ lamella. Good preparation of a device should lead to a large R3 value and small R2 value. In that case, the voltage drop on C and R3 should be the same since the conductive channels can be imagined to be in a parallel geometry to a lamella (depicted as capacitor C). If R3 is large, then the applied potential drop at the lamella C is equivalent to the nominally applied potential (as in schematic +3.5 V). On the other hand, if R3 is small, comparably large current flows through the circuit and most of the potential drops within R1 and R2. Therefore, the voltage on C (and therefore electric field within lamella) is considerably smaller.

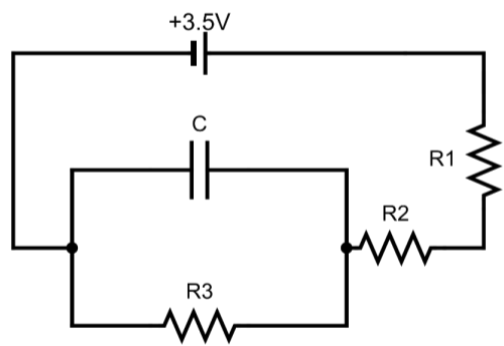

FIG. S2. Equivalent circuit of *in situ* biasing system. R1 represents the resistance of the wires and Pt lines on the chip leading to the lamella. R2 is the resistance of the FIB deposited Pt contacting layers between the $BaTiO_3$ lamella and the Pt electric lines on the chip. C represents the insulating lamella as a capacitor and R3 is equivalent to conducting channels within this lamella.

The FEM simulation is used to explore influence of the sample's geometry on the magnitude and distribution of the electric field within the lamella. It is noted that the calculated electric field values from FEM are of the same order of magnitude but due to leakage effects and non-ideal contact layers they are not expected to match one-to-one. The discrepancy between the FEM calculated electric field magnitude and the one applied during biasing experiments can be largely attributed to the FIB deposited Pt contacting layers between the BaTiO3 lamella and the electric lines of the MEMS chip. In the equivalent circuit above this is represented as R2. During the experiments resistance of the system is 33 MΩ. Comparably R1 is miniscule and can be easily neglected, therefore R2+R3 = 33 MΩ. The configuration of R2 and R3 in the equivalent circuit forms a voltage divider. Voltage drop $U_C$ on C in the equivalent circuit is as follows: $U_C = \frac{R3}{R3+R2} \times 3.5\ V$. If the contacting layers are of good quality R2 is small and the voltage drop on the C is the same as the supplied voltage to the circuit. On the other hand, if R2 ~ R3, the voltage drop across the C is smaller than the supplied voltage. In our case, we expect R3 > R2. That is our sample preparation leads to a situation, where the resistance of the lamella is larger than the resistance of the contacting layers, and this is supported by directly observing the domain response at moderate voltages. Nevertheless, the largest unknown in the system is the quality of the contact between the lamella and the MEMS Pt electric lines, which could improve the *in situ* biasing experiments of the ferroelectric samples.

**Measurement of the domain area – bias loops**

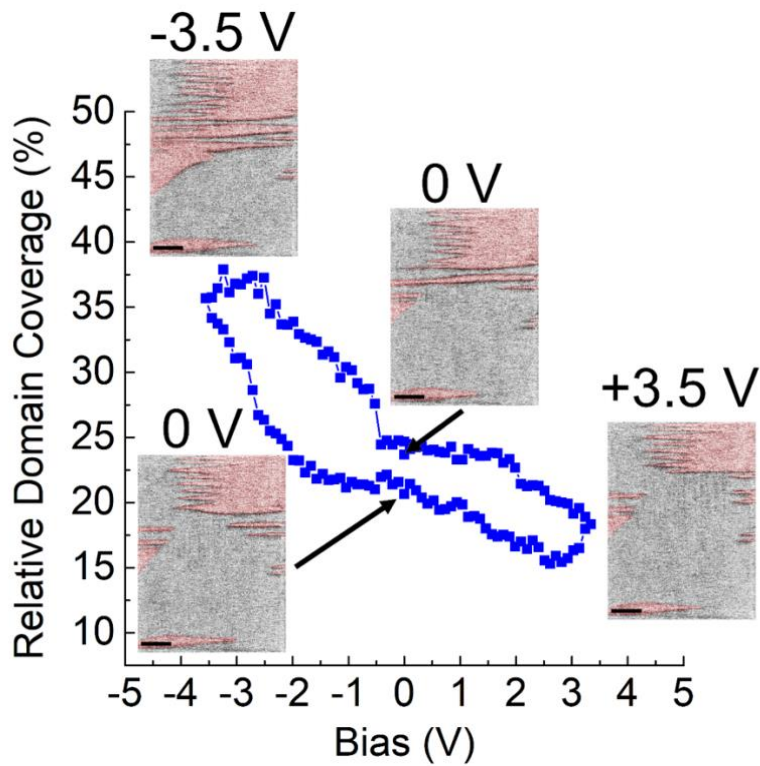

FIG. S3. Domain area – bias loop with images at 0 and ±3.5 V bias voltages (scale bar is 500 nm). Each dot in the loop represents a measurement from one image. Selection of the appropriate domains (red shading in the STEM images) were made based on their response on the electric field – for example, only the domains, which grow with negative bias were measured. Domain area measurements were made manually for each image with ImageJ. The measured domain area was then divided by the total image area, which results in relative domain coverage in the given image.